\documentstyle[epsf,twocolumn,aps,pre]{revtex}

\newcommand{\eg}{\textit{eg.} }

\newcommand{\definedas}{\equiv}

\newcommand{\expectation}[1]{\left\langle #1 \right\rangle}

\newcommand{\greaterthanorabout}
           {\mathrel{\raise.3ex\hbox{$>$\kern-.75em\lower1ex\hbox{$\sim$}}}}
\newcommand{\vol}[1]{{\bf #1}}
\newcommand{\artitle}[1]{}
\long\def\omitt#1{}

\begin{document}
\author{Alexei V. Tkachenko and Thomas A. Witten\\
 \it The James Franck Institute, \\
\it The University of Chicago, Chicago, Illinois 60637 }
\date{to be submitted to  {\sl Phys. Rev. E}}
\title{\bf  Stress in frictionless granular material: Adaptive Network Simulations}
\maketitle
\begin{abstract}
  We present a minimalistic approach to simulations of force transmission
through granular systems. We  start from a configuration containing
cohesive (tensile) contact forces and use an   adaptive procedure  to find the stable configuration with no tensile contact forces. The procedure works by sequentially
removing and adding individual contacts between adjacent beads, while   the bead
positions are not modified.  In a series of two-dimensional
realizations, the resulting force networks are shown to satisfy a linear
constraint among the three components of average stress, as anticipated by
recent theories.  The coefficients in the linear constraint remain
nearly constant for a range of shear loadings up to about .6 of the normal
loading.  The spatial distribution of contact forces shows strong
concentration along ``force chains".  The probability of contact forces of
magnitude
$f$ shows an exponential falloff with $f$.  The response to a local perturbing
force is concentrated along two characteristic rays directed downward and
laterally.

\end{abstract}

\section{Introduction}   \label{Introduction}
The fragility of granular matter is a longstanding preoccupation of engineers
\cite{Coulomb.law,standard.engineering.text} and a recent preoccupation of
physicists\cite{Cates.et.al,Cargese.school}.  By granular matter we mean a
static assembly of hard, spheroidal grains whose contact forces may be
compressive but not tensile.  Thus granular matter is noncohesive.  Mohr and
Coulomb  recognized a fundamental continuum consequence of the noncohesive
state.  There can be no co-ordinate system in which the shear stress exceeds
some fixed multiple $\mu$ of the normal stress
\cite{standard.engineering.text}.  This ``Mohr-Coulomb" condition limits the
stresses that a granular material can support, and thus amounts to a form of
fragility.  When this condition is violated, building foundations settle and
embankments slip. 

Modern civil engineering practice\cite{savage? Goddard?} views the stress field in a granular medium as divided into elastic and plastic zones.  The
stresses in these plastic zones are at the Mohr-Coulomb limit, and thus these zones are at the
margin of stability.  The stress in the elastic zones is within the bounds of stability and
thus the stress here is transmitted as in an elastic body. 

Recently attention has turned to the microscopic origin of the macroscopic fragility of granular media.  The microscopic pattern of contact forces and bead motions shows strong local
heterogeneity and history
dependence\cite{Mueth,Gollub,Novi-sad,Thornton,Coppersmith,Rajchenbach}.   The history of
prior motion in a region clearly influences the way it transmits forces.  The prior motion may
affect the $\mu$ coefficient in the Mohr-Coulomb law, the elasticity tensor, or further
constitutive properties\cite{Cates.et.al}.  The question is, for a given history of
relaxation to a static state, how are these forces transmitted and what range of forces can
be supported.   The transmission of forces can be expressed as a linear-response property of
a granular pack.  An infinitessimal force is added to the bead at position ${\bf x}_0$ and
the corresponding incremental force on a contact at $r$ is determined.  For sufficiently
small perturbations of a finite pack, this linear response function {\bf G} is well defined. 
It depends on the shapes and sizes of the beads, their frictional properties, and how the
pack was constructed. 

 If the perturbing forces become too large, motion occurs.  Beads shift their positions and
form new contacts.  This motion may be reversible, so that the beads return to their original
positions when the perturbation is removed.  This motion may also be irreversible, with the
positions altered after removal of the perturbation.  The thresholds for reversible and for
irreversible motion are fundamental ways to characterize the nonlinear response of the pack. 
For any given pack there is a weakest perturbing force distribution that causes motion.  This
threshold force may go to zero as the size of the pack grows.   Many  simulations have
sought to characterize the above features of a granular
pack\cite{Herrmann.review,Durian.foam,Thornton.simulation}.  These studies model the system
in a realistic way that requires detailed specifications and many parameters.  This detail
makes it difficult to discern which observed features are inescapable consequences of the
granular state, and which are properties of the particular realization.  In this study we
take the opposite approach, sacrificing realism for the sake of simplicity.  We seek the
simplest system that shows the instabilities of noncohesive material.  Thus our system
consists of frictionless, spherical beads, which have been deposited into a container one at
a time and not moved thereafter.  Such a system develops tensile contacts.  

To avoid these contacts, we must define some motion that evolves the pack to a more stable
state.  Again we choose a procedure favoring simplicity rather than realism.  We seek the
stable state attainable with minimal disturbance from the initial state.  Accordingly our
procedure does not move the beads, but rather removes and adds contacts one at a time in
order to attain a stable contact network.  In this sense our simulation is an adaptive
network. 

 \omitt{Since
frictional forces are always absent, this}
 This network is {\it isostatic}\cite{Moukarzel}: the contact forces are determined from
the applied forces solely through the force equilibrium of each bead, without reference to
bead displacements or material deformation.

Our adaptive method demonstrates that frictionless granular materials can be mechanically
robust.  For a given load, the simulation converges \omitt{in polynomial time}  to a state of no tensile contacts.  A change in the applied load of order unity
can be applied with only minor shifts in the contacts.  Further, the three components of the
stress in two dimensions obey a constitutive law of the ``null stress'' type: a weighted sum
of the three components vanishes, the weights depending on the packing but not on the
loading.  

\section{Response Function}

 Our system is a set of spherical beads, whose radii are chosen randomly within a moderate range.  These beads are supported on one side, called the bottom, with a layer of fixed spheres.  The width $w$ of the system is much larger than a bead. 
 \omitt{ (If the system has $d$ dimensions, its size is $w$ in the $d-1$
directions perpendicular to the bottom.)}  The beads are arranged densely in
this space up to a height $h$.  A fixed downward force $F_0$ is applied to each
bead lying at the upper surface.  We choose a configuration of beads that is
mechanically stable: the normal forces acting at the bead contacts oppose the
applied forces $F_0$ and prevent motion. Initially, we allow for tensile
(negative) contact forces.

 We label the $N$ beads by an index $\alpha$.  Then we may denote the contact
force from bead $\alpha$ to bead $\beta$ by the scalar $f_{\alpha\beta}$. 
Newton's Third Law dictates that for all $\alpha$, $\beta$, $f_{\beta \alpha} =
f_{\alpha\beta}$.  The $N_c$ contact forces are constrained by the requirement
that the total vector force on each bead vanish.  In $d$ dimensions, there are
evidently $d~N$ such constraints.  The bead positions ${\bf x}_\alpha$ are
likewise constrained by the geometrical condition that the distance between two
contacting beads $\alpha$ and $\beta$ must be the sum of their radii $r_\alpha
+ r_\beta$.  There are $N_c$ such constraints for the $d~N$ quantities ${\bf
x}_\alpha$.  If all these constraints are independent, the number $N_c$ of
contacts must be exactly $d~N$\cite{Tkachenko.Witten}.  Then the system is
isostatic: the $d~N$ force balance equations are just sufficient to determine
the $N_c$ contact forces.  The equations of force balance may be written
\begin{equation}\label{balance}
\sum_{\beta(\alpha)}f_{\alpha \beta}{\hat n}_{\alpha \beta}={\bf F}_\alpha
\end{equation} 
Here $\beta(\alpha)$ denotes the set of contacting neighbors of  bead  $\alpha$,
${\bf F}_\alpha$ is an external force applied to this bead, $f_{\alpha \beta}$ and  the
unit vector ${\hat n}_{\alpha \beta}$ represent  the magnitude and (fixed) direction of the 
contact force between the two beads $\alpha$ and $\beta$.  Since all the above equations
are linear, the response of the system to a given  external forcing  is determined by the
response function ${\bf G}$:
\begin{equation} 
f_{\alpha \beta} = {\bf G}(\alpha \beta {\big |}  \gamma) \cdot {\bf F}_\gamma
\end{equation} 
\par The  response function ${\bf G}$ determines not only the response to an external
force but also the global displacement field associated with local geometrical perturbation
of the network. In order to see this, let us assume that the packing is subjected to  external
forcing  
${\bf F}_\gamma$.   Then we relax exactly one of the $N$ geometric constraints, and change
infinitesimally the distance between two contacting beads, $r_{\alpha \beta} \equiv |{\bf
x}_\beta -{\bf x}_\alpha|$. As long as the connectivity of the network does not change, its
motion is  non-dissipative. This means that the work done to distort the packing locally,
$\delta r_{\alpha \beta}f_{\alpha \beta}$ is the work against external forces, i. e.

\begin{equation} 
\delta{\bf x}_\gamma \cdot  {\bf F}_\gamma=\delta r_ {\alpha \beta}f_{\alpha \beta}=\delta r_{\alpha \beta}{\bf G}(\alpha \beta {\big |}  \gamma) \cdot {\bf F}_\gamma 
\end{equation} 

The above equation should be valid for any set of external forces; hence,
\begin{equation} \label{displ} 
 \delta{\bf x}_\gamma ={\bf G}(\alpha \beta {\big |}  \gamma)\delta r_ {\alpha \beta}
\end{equation} 
We conclude that ${\bf G}$ is the response function  both for contact force and the displacement field.
Note that the displacement  discussed here is not  due to deformation of the beads.
\omitt{the one associated with the stress (in the limit of very rigid beads the contact force
can be changed without any displacement).}  It  corresponds to
 a ``soft mode'' that preserves the  distances between all contacting
beads other than the  perturbed contact.  

In a general case, finding the response function for a given configuration is a non-local problem, which   requires solving  a set of linear Eq. (\ref{balance}). The task becomes much easier for the case of {\em sequential packing}.  This is created by adding one bead
 at a time. The requirement of mechanical stability implies that  any
newly--added  bead has exactly $d$ ``supporting'' contacts (in $d$-dimensional
space). If all the contacts were permanent and this $d$-branch tree structure
were not perturbed by the future manipulations, the response function might be
found by a simple unidirectional projection procedure. Indeed,
 since there are
exactly $d$ supporting contacts for any bead in a sequential packing,
\omitt{any 
 force (external or applied form the ``supported'' beads)}
 the total force $\tilde {\bf F}_\alpha$ including external force ${\bf
F}_\alpha$ and that applied from the supported beads
can be uniquely decomposed
onto the corresponding
$d$ components, directed along the  \omitt{outcoming}  supporting unit
vectors
${\bf n}_ {\alpha\gamma}$. This gives the values of the  supporting
\omitt{outcoming} forces.  The
$f$'s may be compactly expressed in terms of a generalized scalar product
$\expectation{...|...}_\alpha$:

\begin{equation}
f_{\alpha\beta}= \expectation{\tilde {\bf F}_{\alpha} |{\hat n}_
{\alpha\beta}}_\alpha
\end{equation}

 The scalar product $\expectation{...|...}_\alpha$ is defined such that
$\expectation{{\hat n}_{\alpha \beta}|{\hat n}_
{\alpha\beta'}}_\alpha=\delta_{\beta \beta'}$. 
 for the supporting contacts $\beta$, $\beta'$ of bead $\alpha$.
\omitt{(all the Greek indices
count beads, not spatial dimensions).} In general, it does not coincide with
the 
 conventional
scalar product.      The resulting  response function, ${\bf
G}(\alpha
\beta {\big |} \gamma)$, can be calculated as the superposition of all the
projection sequences (i.e. trajectories), which lead from bead $\gamma$ to the
bond $\alpha\beta$:

\begin{equation}\label{traj}
{\bf G}(\alpha \beta {\big |} \gamma) =\sum_{(\gamma\rightarrow\alpha_1...\rightarrow\alpha\rightarrow\beta)}\left|{\hat
n}_{\gamma\alpha_1}\right\rangle_\gamma \expectation{{\hat
n}_{\gamma\alpha_1}|{\hat
n}_{\alpha_1\alpha_2}}_{\alpha_1}...\expectation{{\hat
n}_{\alpha_k\alpha}|{\hat n}_{\alpha\beta}}_{\alpha}\end{equation}

 Here the summation is  done over all the trajectories
$(\gamma\rightarrow\alpha_1\rightarrow...\rightarrow\alpha_k\rightarrow\alpha\rightarrow\beta)
$ such that any  bead in the sequence is a supporting neighbor of the
previous one.
  
\section{Adaptive network simulation.}

For  a large enough system,  sequential packing is  not compatible with the
requirement of non-tensile contacts.  Anytime when  this  requirement is
violated, a rearrangement occurs and system finds  a ``better'' configuration. One
might expect that this would make  the problem of force propagation a  dynamic 
one. However, it is possible to  limit oneself to  a purely geometrical
consideration, 
 following the ideas of the previous section. 
\omitt{basic ingredient of which have been already introduced in the
previous section.}

Suppose $\alpha\beta$ is a ``bad bond'', 
 whose 
\omitt{i.e. the corresponding} contact force is negative (tensile). This means
that the network would move in such a way that the two beads,   $\alpha$ and
$\beta$ are taken apart. In other words, the  soft \omitt{zero} mode
associated with the perturbation of
$\alpha\beta$ bond is activated, and 
 for small enough displacements
\omitt{ in the linear approximation,}  all the
beads move in accordance with Eq. (\ref{displ}). The motion stops when a
replacement contact is created, i.e. when a gap between any two neighboring
beads   closes. 

In this work, we limit ourselves to this linear approximation. It should be understood that  Eq. (\ref{displ}) is correct only for infinitesimal  displacements, and in a general case one should account for the evolution of the  response function in the course of the rearrangement.  We avoid the problem of changing {\bf G} by permitting only infinitessimal motion in the model.  We imagine that the ``bad bond'' gets deactivated, and it is replaced with a rigid ``strut'' between two neighboring beads that were not in contact in the previous configuration.  
 There is a natural choice for where the strut should be placed to cause minimal disturbance.  Each pair of non-contacting neighbors $\gamma\delta $ has a gap $r_{\gamma\delta} - r_\gamma - r_\delta$.   When the contact $\alpha \beta$ is removed, the distance $r_{\alpha \beta}$ is allowed to change; this change alters the gaps of other neighbors $\gamma \delta$ as specified by Eq. \ref{displ}.  Extrapolating this linear-response equation, motion $\delta r_{\alpha \beta}$ required to close gap $\gamma \delta$ is  given by
\begin{equation}
\delta r_ {\alpha \beta}=\frac{r_{\gamma \delta}-r_{\gamma} -r_{\delta}}{ {\hat
n}_{\gamma \delta}\cdot [{\bf G}(\alpha \beta {\big |}  \gamma)-{\bf
G}(\alpha \beta {\big |}  \delta)]}
\end{equation}
(For many choices of $\gamma \delta$ the required $\delta r_{\alpha \beta}$ is infinite since the $\alpha \beta$ contact has no effect on the $\gamma \delta$ gap.)  
Using this formula, we identify the gap $\alpha' \beta'$ which would require the smallest change of $r_{\alpha \beta}$ in order to close, and we link this pair by a strut. 

After we have found the replacement bond, the modified  response function can be found {\em without solving} the whole set of force balance equations  (\ref{balance})!  We denote the
response function for the initial packing as ${\bf G}_0$.   The new response function {\bf G} must be such that there is no
longer a contact force $f_{\alpha \beta}$.  In general there is such a force in the
initial packing.  However, we may alter this unwanted force by adding an
external force to some other bead.  We choose to add external forces to
beads $\alpha' $ and $\beta'$ that mimic a contact force: the two forces are equal,
opposite and directed along the unit vector $\hat n_{\alpha' \beta'}$ joining them, with a
strength denoted $f_{\alpha' \beta'}$. Our choice of the replacement pair guarantees that the  effect of 
$\alpha'$ or $\beta'$ on
the $\alpha \beta$ contact is non-vanishing.  Then the force on this contact is given by

\begin{eqnarray}\label{falphabeta}
\lefteqn{f_{\alpha \beta} = \sum_\gamma {\bf F}_\gamma \cdot {\bf G}_0(\alpha \beta {\big |} \gamma) }
\\
&\hskip 2cm plus 2cm minus 2cm + f_{\alpha' \beta'} [{\bf G}_0(\alpha \beta {\big |}  \alpha')-{\bf G}_0(\alpha \beta {\big |}  \beta')]\cdot {\hat n}_{\alpha' \beta'}.\nonumber
\end{eqnarray}

We may make this $f_{\alpha \beta}$ vanish by a proper choice of the external force,
$f_{\alpha' \beta'}= \sum_\gamma {\bf F}_\gamma \cdot {\bf G}(\alpha' \beta' | \gamma) $, where ${\bf G}(\alpha' \beta' {\big |} \gamma)$ is the new response function, found by requiring that $f_{\alpha\beta}$ vanish in Eq. (\ref{falphabeta}):
\begin{equation}
{\bf G}(\alpha' \beta' {\big |} \gamma)= -{{\bf G}_0(\alpha \beta {\big |} \gamma)  \over {\hat n}_{\alpha' \beta'}\cdot [{\bf G}_0(\alpha \beta {\big |}  \alpha')-{\bf G}_0(\alpha \beta {\big |}  \beta')]} .
\end{equation}

A contact force on an arbitrary contact is  now determined by a   combination of  external forces ${\bf F}_\gamma$ and the above--determined $f_{\alpha' \beta'}$. This results in the following expression for ${\bf G}( \lambda \mu {\big |} \gamma)$
 ($\lambda \mu$ other than $ \alpha' \beta'$);
  
\begin{eqnarray}
{\bf G}(\lambda \mu {\big |} \gamma) ={\bf G}_0(\lambda \mu {\big |} \gamma)  
\hskip 2in plus 1in minus 2in
\\
 -{\bf G}_0(\alpha \beta {\big |} \gamma){ {\hat n}_{\alpha' \beta'}\cdot [{\bf G}_0(\lambda \mu {\big |}  \alpha')-{\bf G}_0(\lambda \mu{\big |}  \beta')] \over {\hat n}_{\alpha' \beta'}\cdot [{\bf G}_0(\alpha \beta {\big |}  \alpha')-{\bf G}_0(\alpha \beta {\big |}  \beta')]} .\nonumber
\end{eqnarray}

This prescription gives the response of the pack to a wide class of contact
replacements.   The prescription does not require either the
initial or the final state to be stable: it allows tensile contacts.
Using the contact replacement procedure we may investigate the stability of a
pack systematically.  Our algorithm has two major stages: network preparation and its ``mutation'' via the contact replacement scheme. By repeating this adaptive  procedure sufficiently many times, one may hope to get the stable configuration without tensile forces (for a given loading), just like the real system would do. There is a possibility that the  present   geometry--preserving algorithm could not  stabilize the network. For instance if the tangential component of the surface force  is strong enough, it  is expected to initiate a macroscopic avalanche,  as in a sandpile with slope exceeding the critical angle. This class of rearrangements is beyond  the capabilities of our connectivity mutation scheme. This circumstance has even certain advantages: we can determine the critical slope from our simulations as the direction of the surface force at which the algorithm  stops working. It should be emphasized that our  algorithm can easily be modified to incorporate the
change in the network geometry. The major reason why we use the above
geometry-preserving (``strut'')  approximation is its much higher computational
effectiveness.

\section{Simulation details and results}

\subsection{Method}

We begin by creating a  two-dimensional sequential pack of variable-sized
discs by adding them one by one. The studied system has the following parameters: polydispersity 10\% (bead radii  from 1 to 1.1), number of beads, $N$  from 250 to 500 (the major limitation is the computation time). Although there is no gravitational force acting on the beads in these simulations, the statistics of the packing can be varied by changing the ``pseudo-gravity'' direction, ${\hat g}$. Namely,  while adding a bead to the packing we require ${\hat g}$ to be directed between the two supporting contacts.  
Simultaneously, we calculate the response function, by using the
sum-over-trajectories formula, Eq. (\ref{traj}).

\begin{figure}
 \epsfxsize=\hsize $$\epsfbox{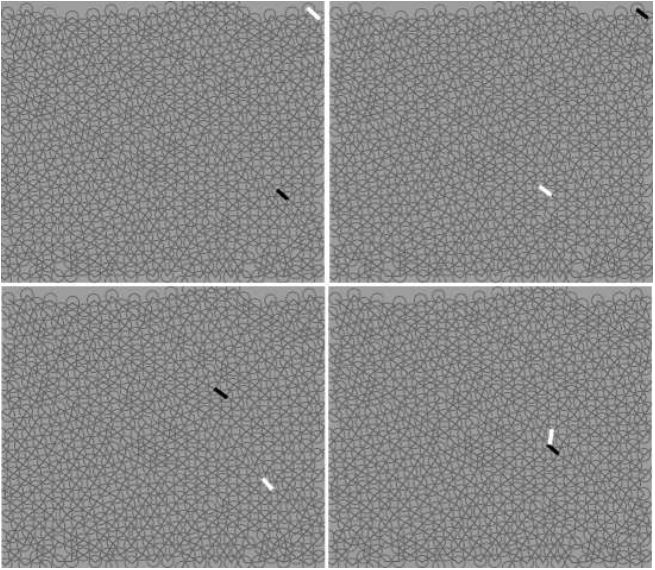}$$
\caption{Four typical steps in our annealing sequence.  The beads are pictured
as circles and the bonds are shown as gray lines.  Upper left to lower
right shows moves 804--807.  The entire annealing process required 1491
such moves.  In each frame the removed contact is shown as a thick white line;
the added contact is shown as a thick black line. }
\label{fourframes}
\end{figure}

Then we apply certain load to  beads on the surface. In the studied cases, the  forces applied to the  surface beads were all the same, with the only principal variable being the ratio of the two components $f_x/f_y$ ($y$ is the vertical direction, the surface in average is  parallel to $x$ due to  the periodic boundary conditions in the horizontal direction).  As long as the response function is given, we   may find all the contact forces for a
given   load.  As we have found,
tensile contacts appear within a few beads from the surface. We analyze the sign of the contact forces one by one, from top to bottom (in the order opposite to the one in which  the beads were originally deposited).   When a tensile
contact is encountered, we follow the contact replacement procedure described in the previous section: find the new bond and modify  the  response function to account for the  connectivity change.  Now we repeat the procedure again starting from very top  until there is no tensile contact left in the
system \omitt{ or as long as practicable (if the algorithm does not
converge).}  Figure \ref{fourframes} shows a sequence of four typical steps in
this procedure.  Evidently the removed and the added contact may be far
apart.

 It sometimes happens that this prescription does not remove the tensile
contacts: the removal of a tensile contact continues to generate others.  In
this case we may modify our procedure for selecting the next tensile contact to
remove.  For example we may select the strongest tensile contact instead of that
tensile contact having the largest sequence number.  Such alternative
prescriptions seem to have little effect on the force network, as discussed
in the next section.

\subsection{ Variability and reproducibility}

While our bond replacement procedure mimics the way  the real system should rearrange, our
choice of the ``bad bond'' to be replaced is far more arbitrary. For instance, instead of
checking the sign of the contact forces one by one from top to bottom, we could go the other
way, or try to replace the contact bearing the largest negative force first. Neither of these
prescriptions is  very realistic;  however, we find that the results are
insensitive to the procedure. \omitt{therefore, our real hope is that the major results of our
simulations are insencitive to this choice.} In order to  probe this sensitivity,
\omitt{check this assumption,} we compared the results from two different ``annealing"
procedures. The first was the procedure described
in  the previous subsection. The second procedure is as follows. \omitt{obtained by using the
algorithm from the previous subsection, with those generated by an alternative, ``annealing''
method. The  idea of the latter is the following.} We perform exactly the same one-by-one
check, as in the previous case, but do not  remove a negative bond unless the  magnitude
of the force exceeds certain tolerance threshold. When no more bonds remain  in the
system for a given  threshold level, we reduce the tolerance and repeat the procedure. The
threshold plays a role of temperature: keeping it finite allows us to deviate from the target
(non-tensile) state of the system and explore its vicinity at the configuration space.
\omitt{There is no surprise that  the} The  second annealing algorithm converges
considerably  faster than the zero-tolerance one. For instance, it took us  from 1500 to 3000
iterations to complete the original algorithm with 500 beads, while the annealing
procedure reduced the needed time to approximately 500-1000 steps.  Interestingly,
\omitt{It is an interesting observation that} the variation of the convergence time is of
order of the time itself. 

\begin{figure}
 \epsfxsize=\hsize $$\epsfbox{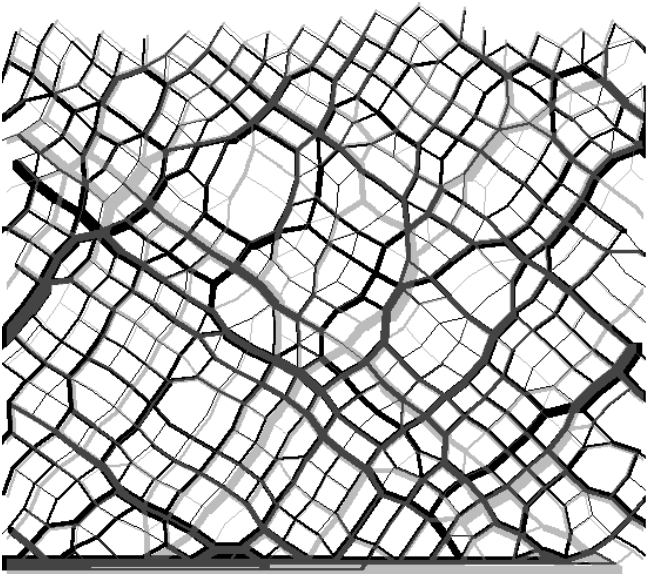}$$
\caption{Pattern of contact forces from two different annealing runs on the
same bead pack.  Each contact force is shown as a line joining the centers of
the two contacting beads.  Thickness of the line is proportional to the
magnitude of the force.  Dark lines show forces obtained by replacing each
tensile contact as it is encountered in a search from the top down.  Light
lines show the forces obtained when only strongly tensile contacts were
replaced at first.  Then when no further strongly tensile contacts remained,
the rest of the tensile contacts were relaxed. The heavy horizontal lines along the bottom
are artifacts introduced by our rendering program.}
\label{chains}
\end{figure}

We have found that the contact configuration resulting from the annealing
procedure does differ from the one generated by the  zero-tolerance algorithm
(see Figure {\ref {chains}}). However, we did not detect any
statistically--significant variation of  the ensemble--averaged properties of
the final state obtained with  the two methods. These properties included   the
average stress and the  contact force probability distribution function (PDF),
presented below. 

\subsection{Macroscopic constitutive equation}

One of the crucial  results of the simulation is that our geometry-preserving  adaptive network algorithm does converge for a considerable range of force direction.  It stops working when $|f_x/f_y|$ approaches $0.6$ (for packing prepared at vertical pseudogravity ${\hat g }$. This suggests that the critical slope for the frictionless packing is about 30 degrees, consistent with simple theoretical arguments and some experiments \cite{angle}
Note that this slope may considerably exceed the angle of repose in dynamic experiments and simulations because of hysteresis associated with the lack of damping in the frictionless system.  Presumably,  the critical slope  can be observed by  quasi-static  tilting of a  zero-slope packing.
    
Another interesting observation is  that the eventual  connectivity of the packing is not too
different from the original sequential packing. For example, the 500-bead system needs up
to  3000 iterative steps (rearrangements) to find the stable state, and yet only 150 out of
1000 contacts ($15 \%$) in the final configuration are different from the original network.
This provides us with a solid background for using the sequential packing as the zero-order
approximation of the real network. This was one of the major hypothesis used in our  earlier
work \cite{Tkachenko.Witten} to derive the constitutive equation of frictionless granular
packing.

One more hypothesis, used for derivation of the macroscopic equation for stress is the mean
field  {\em decoupling Ansatz}.  \omitt{Namely, we have postulated that one can neglect the
correlation between two factors contributing to the microscopic expression for stress:
the force acting at a bead from ``above'', and the third--rank tensor ${\hat \tau}$ associated
with the local contact geometry:}  The average stress in a region of a sequential
packing can be written\cite{Tkachenko.Witten}

\begin{eqnarray} \label{sigma}
\lefteqn{\sigma^{ij}({\bf x})=} 
\\
& \expectation{ \sum_{\alpha}\sum_{\beta(\leftarrow
\alpha)}\delta({\bf x}_\alpha-{\bf x}) \expectation{ {\bf \tilde F}_{\alpha}| {\bf
n}_{\alpha\beta}}_\alpha n_{\alpha\beta}^i n_{\alpha\beta}^j r_{\alpha\beta}},\nonumber
\end{eqnarray}
The sum $\beta$ is over the beads that support the bead $\alpha$.   Our mean-field
hypothesis consists in assuming that the force-related part of this average is independent of
the geometrical part.  We define 
\begin{equation} \label{f}
{\bf f}({\bf x})\definedas\expectation{ \sum_{\alpha}\delta({\bf x}_\alpha-{\bf x})
\langle{\bf \tilde F}_{\alpha}| },
\end{equation}
and 
\begin{equation} \label{tau}
{\hat \tau}\definedas\expectation{\sum_{\beta(\leftarrow
\alpha)}| {\bf n}_{\alpha\beta}\rangle_\alpha
n_{\alpha\beta}^i n_{\alpha\beta}^j r_{\alpha\beta}}.
\end{equation}
Then our mean-field assumption amounts to the statement
\begin{equation} \label{meanfield}
\sigma^{ij} = {\bf f} \cdot \hat \tau
\end{equation}\par

\begin{figure}

 \epsfxsize=\hsize $$\epsfbox{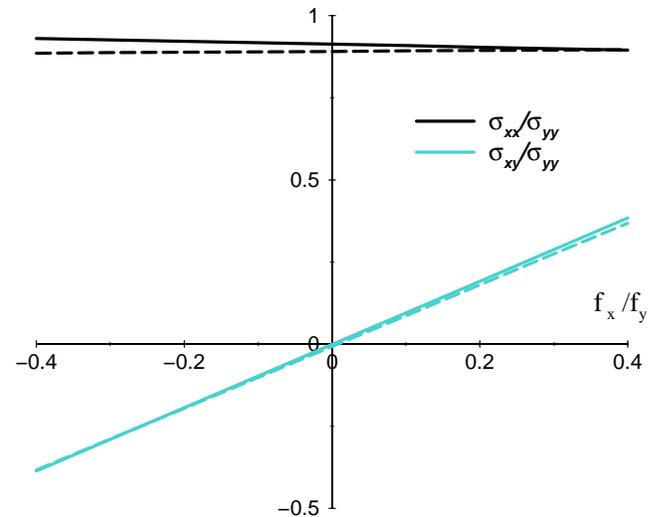}$$
\caption{Various components of stress tensor in original sequential packing (before the adaptive stage), as functions of the direction of the applied force. Note a remarkable agreement between the mean field results (dashed lines) and the simulation data (solid lines). }
\label{stress.before}
\end{figure}

The mean field hypothesis can be directly checked for unperturbed sequential packing, were both fields ${\bf f}$ and ${\hat \tau}$ are well--defined. The results of such a check are represented on Figure {\ref{stress.before}}. The exact values of various stress components are shown to be in an excellent agreement with their evaluation based on the  mean-field Ansatz. We conclude that the mean field is a very good approximation at least for non-adaptive sequential packing.

\begin{figure}

 \epsfxsize=\hsize $$\epsfbox{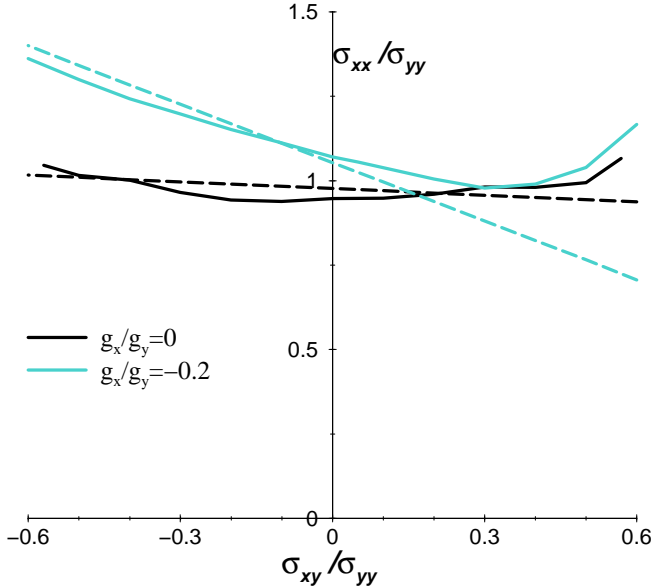}$$
\caption{The only free parameter of the stress tensor, $\sigma_{xx}/\sigma_{yy}$, as a functions of the direction of the applied force, after the adaptive algorithm is completed.  The solid lines show the  simulations results corresponding to two different directions of  ``pseudo-gravity'' vector, $g$.   The dashed lines represent   the null-stress law corresponding to  the  material tensor $\hat\tau$ computed for the original sequential packing. }  
\label{stress.after}
\end{figure}

As long as the rearrangements are switched on, there is no obvious way to define the concept of supporting neighbor, and therefore the ``force from
 subsequent beads'',  
${\bf f}$ is  ill-defined  as well. However, a  more general meaning of   constitutive Eq. (\ref{sigma}), is that the stress is parameterizable with {\em some} vector $\bf f$ , and the third-rank material tensor  ${\hat \tau}$ establishes this parameterization. We now take ${\hat \tau}$ corresponding to the original pre-rearrangement sequential packing and probe the  Ansatz by checking
whether the total stress (after the adaptive procedure) can be expressed as $ {\hat
\tau}\cdot {\bf  f}$. In other words, we compare the only unknown component of the stress
$\sigma_{xx}$ (two other are given by the boundary conditions) with its theoretical value
 obtained from $\hat \tau$. \omitt{following from the  constitutive equation.} We
performed this check for two different classes of packing, corresponding to different
directions of ``pseudogravity'', 
$g_x/g_y=0$ and
$0.2$. The agreement between the two curves  is surprisingly good as long as the direction of
the applied force does not deviate too much from the preparation conditions (i. e. from the
pseudogravity vector), see Figure {\ref{stress.after}}.

\subsection{ Response function}   
\begin{figure}
 \epsfxsize=\hsize $$\epsfbox{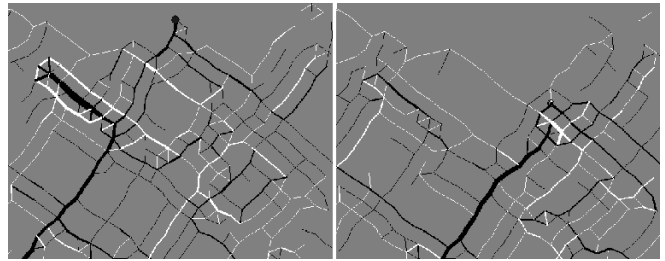}$$
\caption{Two patterns of contact forces resulting from a perturbing point force
in a single relaxed configuration. Perturbation was an infinitessimal downward
force applied at the points shown as heavy dots.  A compressive contact force
is  indicated as a dark line joining the centers of the two contacting beads. 
Thickness is proportional to the magnitude of the force.  Tensile forces are
indicated as lighter gray lines.   }
\label{response.map}
\end{figure}

As noted above, our system transmits forces in accord with a null-stress
constitutive property.  Given the null-stress law, one may infer the
corresponding response function {\bf G}.   The force transmission is transmitted from a
point source according to a wave-like equation\cite{Cates.et.al}.  In a medium where all
non-vertical directions are equivalent, the force should propagate downward
along slanting characteristic lines, whose slope is dictated by the only
parameter  in the null-stress law.  The responding region lies within the
``light cone'' bounded by these lines.  In two dimensions, the response
consists of two delta-functions traveling along the light cone.    Disorder is
expected to scatter the wave solutions of the pure system, thus resulting in a
widening of the delta-peaks. This scattering could be sufficient to create
qualitative new mesoscopic behavior from localization
effects\cite{Anderson?}.   Our simulated system showed strong influences from
disorder, as illustrated in Figure \ref{response.map}.  Because there can be
no vertical-force response at the top of the system, we observe a global
anisotropy, with stronger responses below the source than above it.  The
response is  also strongly heterogeneous. 

\begin{figure}

 \epsfxsize=\hsize $$\epsfbox{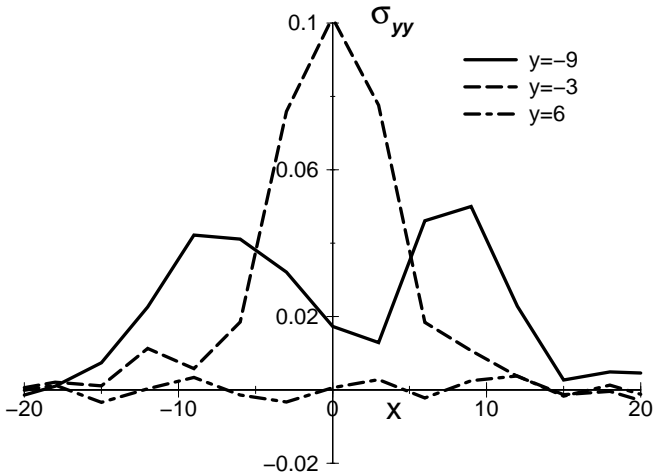}$$
\caption{$\sigma_{yy}$ component of the ensemble--averaged  response to a unit  vertical force applied at the origin $(x=0,y=0)$. The response is measured at several  horizontal cross-sections below ($y=-3,-9$) and above (y=6) the source.} 
\label{response.field}
\end{figure}

  Our simulations allow us to perform ensemble averaging of the response
stress field. Figure \ref{response.field} shows the results of such averaging
over 600 realizations of the network. As the perturbation propagates deep into
the sample, the response function gets a two--peak shape, in a good agreement
with the null-stress law. As \omitt{it was} expected, the peaks are 
broadened by the disorder, and one cannot resolve them  immediately below the
source. Another important observation, which also supports the null-stress
approach, is that the average response is virtually zero above the source. Note
that we have studied only linear response of the system, 
 so that the perturbation did not change the contact network.
\omitt{which assumed  change
in the network geometry.} This need not to be the case in the experiments
involving strong local perturbations \cite{Mueth.private}     


\subsection{Contact force distribution function}

  Our simulation allows us to address yet another interesting and
widely-discussed problem: the statistics of contact force. Recent experiments
indicate that this distribution can be well approximated as exponential, that
is, it is considerably  wider than a naively-expected Gaussian. This is related
to the strong heterogeneities of the mesoscopic  stress in  granular matter: it
appears to be localized to string-like structures known as force chains. 

\begin{figure}
 \epsfxsize=\hsize $$\epsfbox{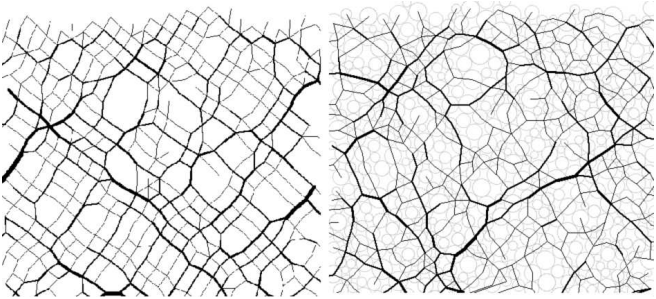}$$
\caption{Pattern of contact forces from two bead packs of different
polydispersity.  Upper picture has ten percent variation in bead radius; lower
picture has three hundred percent variation.  Each contact force is shown as a
line joining the centers of the two contacting beads.  Thickness of the line is
proportional to the magnitude of the force.  In the lower picture the bead
positions are indicated in light gray. }
\label{chains.poly}
\end{figure}

 In the initial sequential packing \omitt{As long as unperturbed sequential packing is
concerned,} there is no constraint on the sign of the contact force, and its amplitude appears
to grow indefinitely with the packing depth. After the rearrangements, there are no
negative forces in the system, and therefore their amplitude cannot grow
forever(the total transmitted force is fixed). Figure  
{\ref{chains.poly}}  shows the spatial distribution of the contact forces in
the systems of two  different degrees of polydispersity  {\em after} the
adaptive stage is completed. One can clearly see that our simulations are at
least in qualitative agreement with experiment: it is easy to identify the
force chains in both cases.

\begin{figure}

 \epsfxsize=\hsize $$\epsfbox{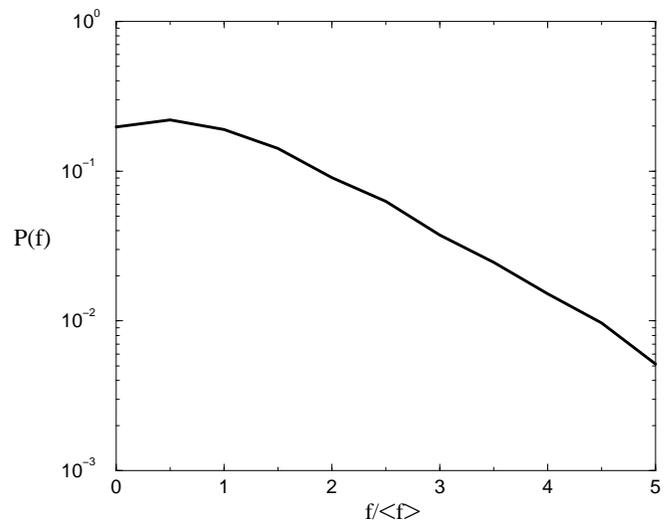}$$
\caption{Probability distribution function of the contact force in granular packing.}
\label{PDF}
\end{figure}

 We were also able to make  a  quantitative comparison between the simulations and experiments. Figure {\ref{PDF}} shows the probability distribution function of the contact force taken from our simulations of almost monodisperse system.   It  apparently  agrees with the exponential  histogram observed experimentally. An insight into the origin of  this  exponential behavior   is given by the   ``q--model'' due to S. Coppersmith {\em et al}\cite{Coppersmith}. The further discussion of this intriguing result will be published elsewhere \cite{Tkachenko.Witten.tobe} 
 
\section{Conclusion}

 In the study of granular materials, clearcut confirmation of theories has been elusive.  One predicted feature of great interest is the null-stress constitutive law postulated by
Wittmer {\em et al}\cite{Cates.et.al}.  We have verified that null-stress
behavior occurs in a simplified granular system embodying disorder, perfect
rigidity and cohesionless contacts.  We have measured the free parameter in the
null-stress law for several situations. 
 We have confirmed  the validity of our major assumptions used for
microscopic foundation of this constitutive law\cite{Tkachenko.Witten}.  Our
simulation also allowed us to compute directly the ensemble--averaged response
function, thus providing an additional check for the  adequacy of the
null-stress approach.

 The simulation method has further interesting features.  It demonstrates that stable configurations of isostatic force networks can be found without changing the positions of the nodes.  It also reveals order-unity variability in the microscopic force distribution resulting from the relaxation process.  Finally, it shows strongly heterogeneous response to point forces---much stronger than that of the geometric contact network.  This suggests strong multiple-scattering features in the force propagation.

The observed exponential  probability distribution function for the contact force is in a good agreement with the experiments. Since there is also an  indirect experimental support for the  null-stress law, our choice of the system (hard frictionless spheres) appears to be an adequate simplification to capture the basic physics of granular rigidity. The further simplifications, such as the fixed--geometry adaptive algorithm provide an effective tool for the future studies of this problem. This may include a study of  non-linear response of the system to large localized perturbation, effects of polydispersity, and history dependence of the response.

  \omitt{
  The question now is whether such features occur more generically in more realistic bead packs.

}

\section*{Acknowledgement}
The authors thank R. Ball, S. Coppersmith, D.  Mueth, H.  Jaeger, S.  Nagel,
and J.Socolar for valuable discussions.  Likewise, we thank the
participants in the Jamming and Rheology program of the Institute of
Theoretical Physics in Autumn 1997.  This work was
supported in part by the National Science Foundation under Award numbers   
PHY-94 07194, DMR-9528957, DMR-9975533 and DMR 94 00379.

\end{document}